\documentclass[acmsmall]{acmart}

\usepackage{tabularx}
\usepackage{caption}
\usepackage{booktabs}
\usepackage{adjustbox}
\usepackage{longtable}
\usepackage{array}
\usepackage{ragged2e}
\newcolumntype{L}[1]{>{\RaggedRight\arraybackslash}p{#1}}
\newcolumntype{Y}{>{\RaggedRight\arraybackslash}X}

\usepackage{geometry}
\usepackage{etoolbox}
\usepackage{makecell}
\AtBeginDocument{%
  }

\setcopyright{none}
\copyrightyear{2018}
\acmYear{2018}
\acmDOI{X.X}
\begin{document}

\title{Generative AI as a Geopolitical Factor in Industry 5.0: Sovereignty, Access, and Control}

\author{Azmine Toushik Wasi}
\orcid{0000-0001-9509-5804}
\affiliation{%
  \institution{Shahjalal University of Science and Technology}
  \city{Sylhet}
  \country{Bangladesh}}
  \email{azmine32@student.sust.edu}

\author{Enjamamul Haque Eram}
\orcid{0009-0005-1088-3702}
\affiliation{%
  \institution{Shahjalal University of Science and Technology}
  \city{Sylhet}
  \country{Bangladesh}}
  \email{enjamamul45@student.sust.edu}
  
\author{Sabrina Afroz Mitu}
\orcid{0009-0007-7456-852X}
\affiliation{%
  \institution{Shahjalal University of Science and Technology}
  \city{Sylhet}
  \country{Bangladesh}}
  \email{sabrina76@student.sust.edu}

  \author{Md Manjurul Ahsan}
\orcid{0000-0003-0900-7930}
\authornote{Corresponding author.} 
\affiliation{%
  \institution{University of OklahomA}
  \city{Oval Norman}
  \country{USA}}
  \email{ahsan@ou.edu}

\renewcommand{\shortauthors}{Wasi et al.}

\begin{abstract}
Industry 5.0 marks a new phase in industrial evolution, emphasizing human-centricity, sustainability, and resilience through the integration of advanced technologies. Within this evolving landscape, Generative AI (GenAI) and autonomous systems are not only transforming industrial processes but also emerging as pivotal geopolitical instruments. We examine strategic implications of GenAI in Industry 5.0, arguing that these technologies have become national assets central to sovereignty, access, and global influence. As countries compete for AI supremacy, growing disparities in talent, computational infrastructure, and data access are reshaping global power hierarchies and accelerating the fragmentation of the digital economy. The human-centric ethos of Industry 5.0, anchored in collaboration between humans and intelligent systems, increasingly conflicts with the autonomy and opacity of GenAI, raising urgent governance challenges related to meaningful human control, dual-use risks, and accountability. We analyze how these dynamics influence defense strategies, industrial competitiveness, and supply chain resilience, including the geopolitical weaponization of export controls and the rise of data sovereignty. Our contribution synthesizes technological, economic, and ethical perspectives to propose a comprehensive framework for navigating the intersection of GenAI and geopolitics. We call for governance models that balance national autonomy with international coordination while safeguarding human-centric values in an increasingly AI-driven world.
\end{abstract}


\settopmatter{printacmref=false}
\maketitle


\section{Introduction}
Rise of Generative Artificial Intelligence (GenAI) and the spread of autonomous systems are bringing about a major transformation. These technologies are not just improving how things work, rather they are reshaping entire industries \cite{103} and changing the balance of power between nations. This paper argues that GenAI and autonomy have moved beyond being just cutting-edge tools. Today, they are powerful strategic assets that affect how strong a country’s supply chains are and how well its industries can compete globally. Because of this growing influence, there is an urgent need to create new rules and systems to manage them properly. This change is happening during the shift toward Industry 5.0, creating a new phase of industrial development that focuses on putting people back at the center. Unlike previous stages, which prioritized automation and efficiency, Industry 5.0 values human creativity, ethical responsibility, and collaboration between humans and machines. However, this creates a unique tension. As GenAI becomes more independent and capable of making decisions on its own, it raises important questions about how to balance automation with human oversight and ensure these systems serve people, not the other way around.


Generative AI (GenAI) represents a sophisticated advancement in artificial intelligence, distinguished by its ability to autonomously generate novel content, including images, videos, text, and complex designs, thereby extending far beyond traditional AI’s focus on pattern recognition and decision support  \cite{1}. The evolution of GenAI from early models such as Variational Autoencoders (VAEs) to more advanced architectures like Generative Adversarial Networks (GANs) has catalyzed significant transformation across industrial and commercial sectors  \cite{1}. These capabilities have been further amplified by the democratization of access to Large Language Models (LLMs), which are trained on massive datasets and now widely deployed to generate high-quality, human-like outputs  \cite{2}. In parallel, autonomous systems, often defined as intelligent, self-directed entities that integrate mechanical systems, AI, sensors, and connectivity—have matured into operational tools capable of executing complex tasks in dynamic physical environments with minimal or no human oversight  \cite{3}. Such systems span a broad array of applications, from autonomous vehicles and drones to intelligent industrial robotics  \cite{3}. Their deployment within smart manufacturing environments demonstrates how they can integrate human adaptability, machine precision, and advanced computational capacity to establish highly resilient and sustainable production architectures  \cite{101}. These developments are converging within the broader transition to Industry 5.0, a paradigm that marks a strategic shift from the automation-driven ethos of Industry 4.0 toward a model that re-centers human agency, creativity, and values  \cite{1}. Unlike its predecessor, Industry 5.0 aims to reframe industrial progress by embedding sustainability, resilience, and ethical responsibility as core design principles, alongside the adoption of advanced technologies such as GenAI and robotics  \cite{8}. Within this emergent framework, GenAI is proving instrumental in advancing critical industrial functions including customer service automation, quality assurance, process optimization, and predictive maintenance  \cite{7}. However, the role of GenAI and autonomous systems in Industry 5.0 is not framed around substitution of human labor, but rather its augmentation. This is reflected in the growing emphasis on human-AI collaboration, wherein AI handles repetitive or hazardous tasks while enhancing human capacity for complex reasoning, creativity, and problem-solving  \cite{6}. As such, the strategic advantage in Industry 5.0 lies not merely in technology adoption but in fostering synergistic human-AI relationships that preserve human judgment at the center of innovation. Nations and organizations that successfully navigate this symbiosis are likely to establish leadership positions in the evolving industrial order, characterized by enhanced flexibility, robustness, and social accountability  \cite{6}. This shift underscores a broader reimagining of productivity and progress, in which human well-being and ethical considerations are no longer peripheral, but foundational to industrial advancement.


Industry 5.0 marks a significant departure from the automation-driven ethos of Industry 4.0, embracing a more holistic vision that places human centrality, sustainability, and resilience at the forefront of manufacturing paradigms  \cite{7}. This shift envisions a synergistic collaboration between human workers and advanced technologies—such as collaborative robots (cobots), artificial intelligence (AI), and intelligent automation systems—working together to facilitate highly customized, efficient, and environmentally responsible production processes  \cite{6}. Central to enabling this transformation are technologies that not only enhance productivity but are deliberately designed to empower and complement human capabilities. Cobots, for instance, are engineered to operate safely alongside workers, facilitating shared tasks in dynamic settings  \cite{13}. AI systems contribute by enabling real-time rescheduling and adaptive flow control, while digital twins offer comprehensive simulation and testing environments to optimize operations before implementation  \cite{13}. Augmented Reality (AR) enhances human-machine interfaces by delivering real-time, contextual instructions to operators, and edge computing ensures low-latency responsiveness, enabling fluid human-machine interaction  \cite{13}. Collectively, these technologies exemplify the Industry 5.0 principle of making machines \textit{worth working with} rather than tools of substitution  \cite{13}. The advantages of this human-centric approach are considerable, including enhanced product personalization, improved sustainability through intelligent resource management, and higher levels of worker satisfaction and safety  \cite{6}. By reintroducing human creativity, judgment, and craftsmanship into manufacturing processes, Industry 5.0 shifts labor from routine and hazardous tasks to roles focused on supervision, design, and innovation  \cite{6}. However, this vision is complicated by the accelerating capabilities of Generative AI (GenAI), which increasingly exhibits agentic behaviors such as autonomous decision-making and self-directed task execution  \cite{5}. Unlike earlier systems, these AI agents can function with minimal human oversight and often operate as opaque \textit{black boxes,} raising concerns about transparency and accountability in decision-making  \cite{5}. This dynamic introduces a critical governance dilemma: how to retain \textit{meaningful human control} over systems that are becoming more independent, especially in high-stakes contexts where errors could have severe ethical or legal implications  \cite{16}. The potential for AI to implicitly assume moral or legal responsibilities necessitates not only technical safeguards but also adaptive regulatory frameworks that clearly delineate accountability between humans and machines  \cite{11}. As such, realizing the promises of Industry 5.0 will require a delicate balance—leveraging the power of autonomous technologies while ensuring that human agency, values, and responsibility remain central to industrial decision-making.

Generative AI (GenAI) and autonomous systems have transcended their origins as frontier technologies to become core strategic assets, with far-reaching implications for global power dynamics and national sovereignty  \cite{17}. Much like energy resources or military capabilities, these technologies now serve as foundational determinants of geopolitical influence, placing them at the center of international competition and strategic policymaking. Their strategic significance is particularly evident in the domain of national security, where AI is revolutionizing defense architectures by enhancing situational awareness, enabling real-time decision-making, and optimizing the deployment of unmanned systems in complex operational environments  \cite{2}. As a result, nations are making large-scale investments in AI not merely for economic growth, but to safeguard digital sovereignty and bolster national defense infrastructures  \cite{20}. Economically, AI has emerged as a key driver of global asymmetries, fueling a high-stakes race among technologically advanced countries to secure leadership in innovation, talent acquisition, and infrastructure development  \cite{21}. The competitive edge gained through mastery of AI technologies confers not only military superiority but also outsized economic and political leverage on the global stage  \cite{23}. Importantly, this AI arms race extends well beyond conventional military applications, evolving into a broader struggle for influence over global norms, regulatory frameworks, and sociopolitical narratives  \cite{104}. Nations are thus competing for the ability to shape the rules of the digital order and to influence societal outcomes through AI governance, platform control, and data dominance  \cite{22}. In this context, a country's status as a leader in GenAI is defined by multiple factors—its research output, technological ecosystem, workforce capabilities, and ability to translate scientific breakthroughs into scalable, real-world applications  \cite{17}. However, this growing centrality of GenAI also introduces the enduring challenge of dual-use capabilities. The same technologies that enable breakthroughs in healthcare, energy efficiency, and humanitarian response can be repurposed for harmful applications such as autonomous weapons, cyber warfare, disinformation campaigns, and authoritarian surveillance  \cite{7}. This dual-use dilemma complicates governance efforts, as overly restrictive policies could stifle beneficial innovation, while permissive environments risk enabling malign actors and systemic harm  \cite{32}. As such, policymakers face a difficult balancing act: they must reconcile national security interests with economic competitiveness, democratic accountability, and ethical responsibility in the governance of GenAI and autonomous systems.

In this paper, we explore how GenAI is emerging as a key geopolitical factor within the broader framework of Industry 5.0, focusing on the critical themes of sovereignty, access, and control. We examine the evolving role of GenAI and autonomous systems in reshaping national power structures and influencing strategic positioning on the global stage. By adopting a multidisciplinary approach, we analyze how these technologies are transforming defense capabilities, disrupting global supply chain resilience, and redefining the basis of industrial competitiveness. Our analysis highlights the dual-use nature of GenAI, capable of enabling both profound innovation and considerable security challenges. We emphasize that strategic control over GenAI infrastructure, skilled talent, and high-quality data will become central to national influence and digital sovereignty. The global race for AI supremacy is already contributing to the formation of AI blocs and the fragmentation of shared technological standards. Through the synthesis of developments in policy, defense, and economic strategy, we provide a comprehensive perspective on how AI is driving a new phase of geopolitical realignment and recommend policies to handle these issues. Our work addresses the growing tension between autonomous AI capabilities and the human-centric ethos of Industry 5.0. We further explore how this tension creates governance gaps that must be addressed through adaptive and inclusive frameworks. Ultimately, we propose a path forward for responsible governance that aligns technological advancement with global equity, collaboration, and ethical oversight.

\section{Generative AI and Autonomous Systems as Strategic National Assets}
Global landscape is witnessing an intensified competition among nations to establish preeminence in GenAI and autonomous systems. These technologies are increasingly recognized as foundational elements of national power, driving strategic investments and shaping defense capabilities.

\subsection{National AI Strategies and Global Investment Landscape}
As AI technologies rapidly advance, nations worldwide are adopting varied strategies to secure leadership in this critical domain, as outlined in Table \ref{tab:national-ai-strategies}. These approaches reflect diverse political, economic, and security priorities, shaping a complex global investment landscape. Understanding these strategies is essential to grasp the evolving geopolitical dynamics of AI innovation and control.

\subsubsection{Diverse National Strategies and Key Players}
The global race for AI leadership is marked by a variety of national strategies aimed at leveraging AI for economic growth, national security, and digital sovereignty  \cite{18}. These strategies reflect differing political systems, economic goals, and geopolitical ambitions, leading to the emergence of distinct AI innovation blocs and strategic alliances  \cite{22}. The United States follows a largely decentralized, market-driven approach, prioritizing long-term research and development, workforce development, and the establishment of technical standards to promote AI safety and security  \cite{34}. Its strategy relies heavily on a dynamic commercial ecosystem and a steady pipeline of elite AI talent  \cite{18}. The National AI Initiative Act of 2020 formalized these efforts by creating a dedicated office focused on fostering innovation, industry collaboration, workforce readiness, and adherence to fundamental societal values  \cite{34}. In contrast, China pursues a government-led AI strategy with a clear goal to become the world leader by 2030  \cite{20}. It leads in AI research volume, citations, and patent filings, outpacing the United States in many key metrics  \cite{17}. China’s \textit{New Generation AI Plan} coordinates extensive state-backed investments across ministries, fostering rapid AI development despite challenges like U.S. chip restrictions, as demonstrated by projects like the \textit{DeepSeek} Large Language Model  \cite{17,20}.

\subsubsection{European Union and Middle Eastern AI Strategies}
The European Union emphasizes ethical AI development alongside substantial financial investment, targeting approximately EUR 20 billion annually  \cite{37}. The EU fosters collaboration across member states, prioritizes talent retention, and promotes the development of common data spaces and high-performance computing infrastructure  \cite{37}. A central feature of the EU’s approach is the development of comprehensive ethical guidelines and human-centric AI principles  \cite{16}. The EU AI Act represents a landmark regulatory framework, emphasizing transparency, risk mitigation, and rigorous oversight to ensure trustworthy AI deployment  \cite{33}. Meanwhile, Middle Eastern countries, notably the UAE and Saudi Arabia, have positioned AI as a key pillar of economic diversification  \cite{20}. The UAE was an early mover with the world’s first Ministry of AI, aiming for global leadership by 2031 and projecting AI to contribute up to 14\% of its GDP by 2030  \cite{20}. Saudi Arabia’s Vision 2030 similarly commits substantial resources to AI, establishing the Saudi Data \& AI Authority (SDAIA) and rolling out a comprehensive National Strategy for Data \& AI  \cite{20}.

\begin{table}[t]
\centering
\caption{Overview of National AI Governance Frameworks}
\label{tab:national-ai-strategies}
\scalebox{0.85}{
\begin{tabular}{|p{2cm}|p{4cm}|p{3cm}|p{5cm}|}
\hline
\textbf{Country/ Bloc} & \textbf{Key Strategic Focus} & \textbf{Investment Levels (Approx.)} & \textbf{Noteworthy Initiatives/ Characteristics} \\
\hline
United States & Market-driven innovation, R\&D leadership, national security & $\sim$\$109B (private, 2024), significant federal R\&D & Vibrant commercial ecosystem, elite talent, National AI Initiative Act of 2020 \\
\hline
China & State-led AI leadership by 2030, digital sovereignty & $\sim$\$9B (venture, 2024), major state spending & Leading in AI research volume, patents, New Generation AI Plan, DeepSeek LLM \\
\hline
European Union & Human-centric, ethical AI, regulatory leadership & EUR 20B/year (public \& private) & EU AI Act, Common European Data Spaces, AI Gigafactories \\
\hline
UAE & Economic diversification, global AI leadership by 2031 & US\$10B government fund (G42) & First Ministry of AI, homegrown LLMs (Falcon), AI Campus \\
\hline
Saudi Arabia & Economic diversification, world AI leader by 2030 & Large investments in infrastructure & SDAIA, Vision 2030, National Strategy for Data \& AI \\
\hline
Japan & Innovation promotion, soft-law governance & Billions in research and startups & Society 5.0, AI Guidelines for Business \\
\hline
South Korea & Sovereign AI, homegrown models, top 3 AI leader & 100 trillion won (\$73B) investment target & AI National Strategy 2025, strong STEM education \\
\hline
India & AI talent pool, agricultural innovation & Emerging public-private investments & National AI Strategy (NITI Aayog) \\
\hline
Brazil & Agricultural AI innovation & Emerging government efforts & Brazilian AI Strategy \\
\hline
Canada & AI alignment and safety research & Strong academic investment & Pan-Canadian AI Strategy, leader in AI ethics \\
\hline
\end{tabular}}
\end{table}

\subsubsection{Emerging AI Hubs: Asia and Beyond}
Other major players in the AI landscape include Japan, South Korea, India, and Brazil, each advancing distinctive strategies aligned with their national priorities  \cite{18}. Japan’s AI governance policy adopts a \textit{soft-law} approach that emphasizes innovation and international interoperability, consistent with its Society 5.0 vision that integrates technology with societal well-being  \cite{18}. South Korea aims to establish itself as an \textit{AI sovereign} by investing heavily in homegrown AI models, talent development, and startup ecosystems, aspiring to be among the top three AI leaders globally  \cite{18}. India, with a large pool of AI professionals, and Brazil, which is innovating notably in agricultural AI applications, are also important contributors to the global AI ecosystem  \cite{18}. These diverse national strategies illustrate how AI development is becoming a complex and multipolar field, with different countries carving out niches based on their strengths and economic priorities.

\subsubsection{Talent Competition and Geopolitical Fragmentation}
A critical and often underappreciated dimension of AI geopolitics is the intensifying \textit{talent war} for skilled researchers, engineers, and data scientists  \cite{25}. Success in AI is not solely dependent on hardware or data but hinges on human capital capable of designing, training, and deploying sophisticated AI systems. The United States maintains a competitive edge through elite talent pipelines and workforce development initiatives  \cite{34,18}. China boasts the largest AI talent pool globally, supported by extensive government programs and education investments  \cite{17}. South Korea is aggressively expanding its AI talent base and actively seeks to attract foreign professionals to boost its capabilities  \cite{42}. This global competition for talent directly influences national competitiveness and strategic autonomy. Developing countries that fail to retain or attract AI expertise risk widening existing inequalities, as the economic and geopolitical benefits of AI increasingly concentrate in nations with strong talent ecosystems  \cite{47}. These trends contribute to geopolitical fragmentation, with AI innovation and governance becoming increasingly segmented along national and regional lines  \cite{22}.

\subsection{Defense and Security Applications}
Generative AI and autonomous systems are instigating a profound paradigm shift in military research and application, fundamentally enhancing operational efficiency and decision-making processes across defense sectors \cite{2}. These technologies are becoming increasingly critical for a wide array of military and defense applications, including air surveillance systems, missile defense, and cybersecurity \cite{4}. GenAI can detect attack patterns with 87\% accuracy \cite{105}. By strengthening current security frameworks through continuous monitoring, strategies should incorporate GenAI-powered approaches that are sophisticated and difficult to predict \cite{106}.

Several key use cases illustrate the transformative impact of GenAI in defense:

\begin{itemize}
  \item \textbf{Intelligence Summarization and Threat Analysis:} Modern military operations generate immense volumes of data from diverse sources, such as satellite imagery, intercepted communications, and open-source intelligence. GenAI models can automate the summarization of this information, extracting crucial insights, identifying patterns and anomalies indicative of emerging threats, and adapting dynamically to evolving adversary tactics \cite{19}. This capability allows military intelligence units to anticipate and stay ahead of potential threats, providing timely warnings and recommendations.

  \item \textbf{Mission Planning and Simulation:} GenAI assists in complex military mission planning by rapidly generating multiple courses of action, simulating potential outcomes, and identifying optimal strategies. This enhances the agility and responsiveness of military operations, particularly in rapidly evolving conflict zones \cite{19}. For instance, the Pentagon's \textit{Thunderforge} project specifically aims to enhance military planning through AI tools developed in collaboration with tech companies, integrating data from intelligence sources and battlefield sensors to provide commanders with strategic recommendations \cite{19}.

  \item \textbf{Autonomous Drone Coordination:} Ilachinski (2017) defines automated systems as physical systems that operate with minimal human input in stable settings, performing predefined tasks based on fixed rules or scripts \cite{107}. The deployment of autonomous drones in military operations for surveillance, reconnaissance, and combat is being significantly enhanced by GenAI. These AI systems enable real-time decision-making and coordination among drones without direct human intervention. This allows drones to adapt to changing environments, identify targets, and coordinate effectively in swarm operations, thereby enhancing operational efficiency and reducing risk to human personnel in hostile environments \cite{19}.

  \item \textbf{Electronic Warfare Simulation:} GenAI can simulate intricate electronic warfare (EW) scenarios, generating synthetic signals and interference patterns to test and improve defense systems. This capability allows military units to train for and adapt to various EW threats without the need for costly and risky live exercises, facilitating the continuous development of countermeasures and refinement of tactics aGenAInst electronic attacks \cite{19}.
\end{itemize}

Despite significant advancements in integrating AI into intelligence analysis, concerns persist about the reliability and potential biases of AI-generated insights, emphasizing the essential role of human oversight to validate these findings \cite{19}. A key ethical issue is ensuring \textit{meaningful human control and judgment} over critical, life-and-death decisions, as moral and legal responsibilities must not be delegated to machines or software \cite{16}. The increasing deployment of lethal autonomous weapon systems (LAWS) that operate without human intervention has intensified calls for binding international regulations to mitigate associated risks \cite{31}. Additionally, GenAI accelerates military decision cycles by rapidly generating options and simulating outcomes, compressing the time available for human deliberation and conflict de-escalation \cite{2,19}. While this capability enhances situational awareness, it also raises the danger of decisions being made faster than human cognitive capacities can process, potentially triggering unintended escalation or conflicts. This situation highlights the urgent need for stringent technical safeguards and robust human control frameworks to manage the inherent risks of autonomous military AI applications \cite{16}.

\subsection{Race for AI Supremacy and Technological Independence}
Global competition for AI supremacy is intensifying as nations recognize AI’s critical role in shaping future economic, military, and geopolitical power. Achieving technological independence has become a strategic priority to reduce vulnerabilities and secure long-term national influence in the AI-driven world.

\subsubsection{Global Competition for AI Supremacy}
Global race for AI supremacy has become a central aspect of modern geopolitics, driving power imbalances among nations \cite{21}. Countries are actively competing to influence the future development and deployment of AI technologies to secure their strategic positions in the emerging global order \cite{22}. This competition extends beyond mere technological innovation; it involves shaping international standards, controlling critical supply chains, and establishing dominance in AI-related industries \cite{21}. The stakes are high, as AI supremacy translates into military, economic, and political advantages on the global stage. Many nations are also investing heavily in AI research, talent development, and infrastructure to build competitive ecosystems that can sustain long-term innovation \cite{18}. These efforts highlight the urgency and strategic importance placed on AI across a broad spectrum of national priorities \cite{22}.

\subsubsection{Pursuit of Technological Independence}
A key goal for many nations is to achieve technological independence in AI, reducing reliance on foreign technologies and avoiding vulnerabilities associated with external dependencies \cite{42}. South Korea’s explicit strategy of pursuing \textit{sovereign AI} involves developing domestic AI models and nurturing local ecosystems to maintain autonomy \cite{42}. China’s DeepSeek Large Language Model showcases its capacity to innovate despite external chip sanctions, underscoring its drive toward self-sufficiency \cite{17}. Similarly, Gulf states are dedicating substantial resources to build indigenous AI capabilities as part of their broader economic diversification plans \cite{20}. Success in this area depends not only on technological advances but also on the integration of ecosystems, regulatory agility, and resilient supply chains that span national borders \cite{18}. Countries that prioritize coordination, resilience, and specialization are better positioned to shape their technological futures and exert influence internationally \cite{18}.

\subsubsection{Challenges for Developing Nations}
Despite the ambitions of many countries, the pursuit of AI supremacy and independence poses significant challenges, especially for developing nations \cite{21}. Many of these countries risk being marginalized in the AI race due to insufficient infrastructure, limited access to talent, and financial constraints \cite{21}. As a result, they often advocate for enhanced international cooperation and capacity-building initiatives to safeguard their autonomy and participation in the AI ecosystem \cite{21}. Developing nations face the dilemma of balancing the adoption of foreign AI technologies with concerns about security and long-term dependence \cite{21}. Without equitable access to AI resources and governance influence, these countries may become passive consumers rather than active innovators \cite{21}. This imbalance threatens to exacerbate existing global inequalities and undermine collective progress toward inclusive technological development \cite{21}.

\subsubsection{Data Sovereignty and AI Colonialism Risks}
The intertwined nature of AI technological independence and data sovereignty is increasingly shaping geopolitical dynamics \cite{38}. Access to diverse, high-quality datasets is vital for AI innovation, making data a critical strategic asset regulated by national policies \cite{38}. This has led to the rise of data localization policies, where countries seek to retain control over their data within their borders \cite{51}. While data sovereignty strengthens national control, it also risks fragmenting the global digital economy by creating \textit{data borders} that impede the free flow of information essential for AI progress \cite{51}. A significant emerging concern is \textit{AI colonialism,} where powerful nations leverage AI development to reinforce economic and technological dominance over developing countries \cite{47}. High-income countries benefit from superior infrastructure, abundant resources, and advanced data ecosystems \cite{48}, while developing countries may suffer from exploitation through labor or data extraction without fair returns \cite{21}. This dynamic can entrench existing inequalities, marginalize alternative perspectives, and limit low-income countries’ participation in setting global AI norms \cite{27,55}. If unaddressed, these trends threaten to concentrate AI-related wealth and power further, hindering sustainable and equitable global development \cite{48}.

\section{Impact on Global Supply Chain Resilience}
Generative AI and autonomous systems are fundamentally reshaping global supply chains, presenting both unprecedented opportunities for optimization and new, complex vulnerabilities that carry significant geopolitical risks. The influence of data sovereignty further complicates this evolving landscape.

\subsection{Opportunities for GenAI-Driven Optimization} 
Generative AI and broader AI applications are transforming supply chain management by optimizing procedures, enhancing decision-making, and significantly improving overall efficiency \cite{57}. Organizations that strategically leverage AI are positioned to build digital, adaptive, and ultimately autonomous supply chains, thereby GenAIning a distinct competitive advantage in a world that increasingly values efficiency, sustainability, and intelligent decision-making \cite{58}.

Key applications and their benefits include:

\begin{itemize}
  \item \textbf{Predictive Maintenance and Resource Optimization:} GenAI can generate new maintenance plans by correlating with the likely failure times of equipment, reducing costly downtime \cite{59}. For instance, Maersk has decreased vessel downtime by 30\% through AI-driven predictive maintenance, leading to substantial annual savings and reduced carbon emissions \cite{61}. AI also optimizes resource consumption, contributing to sustainability goals \cite{6}. Predictive analytics are crucial for accurately forecasting demand, optimizing supply, and mitigating disruptions before they occur \cite{57}.

  \item \textbf{Demand Forecasting and Inventory Management:} GenAI analyzes vast historical data, market trends, and customer behavior to forecast demand with remarkable accuracy, leading to optimized inventory levels and reduced waste \cite{60}. Walmart, for example, has reduced inventory costs by \$1.5 billion annually through AI inventory management systems, while Nestlé has transformed its inventory forecasting from manual to AI-driven analytics \cite{61}. The deep learning-based model developed in this study greatly improves retailers’ accuracy in forecasting future sales, supporting more informed decisions and effective strategic planning \cite{108}. Incorporating AI into demand forecasting gives manufacturers a strategic advantage by delivering insights into market trends, customer behavior, and sales patterns \cite{111}.

  \item \textbf{Logistics Automation and Optimization:} AI models are instrumental in optimizing freight scheduling, last-mile delivery, and real-time tracking \cite{64}. UPS's ORION route optimization system, powered by AI, processes thousands of route optimizations per minute, saving millions of liters of fuel and reducing carbon emissions \cite{61}. Autonomous vehicles and robots are also enhancing efficiency in warehouses by managing and processing products, and improving the transportation of goods \cite{65}.

  \item \textbf{Risk Management and Scenario Planning:} GenAI can detect and simulate potential disruptions by analyzing complex factors such as congested ports, shipment routes, and vendor mapping. This enables supply chain managers to proactively deploy mitigation strategies and develop robust contingency plans \cite{59}. GenAI's ability to evaluate various options without manual report generation significantly enhances decision-making and overall supply chain resilience \cite{64}.

  \item \textbf{Quality Control:} Computer vision systems, often powered by GenAI, enable automated visual inspection, detecting product defects in real-time and reducing reliance on manual checks \cite{60}. Ford Motor, for instance, uses AI to identify subtle flaws like wrinkles in car seats, improving product quality \cite{63}. Bosch utilizes Generative AI to predict potential defects in automotive parts \cite{109}. Generative AI enables the customization of quality control by creating specific inspection plans tailored to various product types.

  \item \textbf{Network Design and Collaboration:} GenAI can identify optimal network configurations by considering factors such as facility locations, transportation routes, and resource allocation. It generates and evaluates various potential network designs to identify the most efficient arrangement, balancing cost, availability, and risk \cite{59}.
\end{itemize}

These AI-driven capabilities collectively foster enhanced agility and responsiveness in supply chains. They enable businesses to adapt swiftly to changing market conditions and capitalize on emerging opportunities, making supply chains more flexible and resilient, particularly during disruptive events \cite{57}.

The shift from reactive to proactive supply chain management via GenAI is a transformative development. Historically, supply chains have often been reactive to disruptions, responding to crises as they emerge \cite{66}. However, GenAI's capacity to anticipate risks, recommend mitigation actions, and proactively deploy strategies fundamentally shifts this paradigm \cite{59}. This transformation, driven by GenAI's ability to process vast datasets and simulate complex scenarios, allows organizations to \textit{future-proof their supply chains} and build higher levels of resilience \cite{58}. This proactive capability becomes a strategic asset for nations and companies, enabling them to better withstand geopolitical shocks, natural disasters, or economic shifts, ensuring continuity and stability in global trade \cite{66}.

\begin{table}[t]
\caption{Generative AI’s Impact on Supply Chain Functions}
\label{tab:genai-supply}
\scalebox{0.85}{
\begin{tabular}{|p{2cm}|p{4cm}|p{4cm}|p{4cm}|}
\hline
\textbf{Supply Chain Function} & \textbf{GenAI Application/Capability} & \textbf{Key Benefits} & \textbf{Illustrative Examples/Case Studies} \\
\hline
Demand Forecasting & Predictive analytics, pattern recognition from diverse data (sales, trends, customer behavior) & Highly accurate demand prediction, reduced waste, optimized production & Walmart (reduced inventory costs by \$1.5B/yr), Nestlé (transformed forecasting) \cite{30} \\
\hline
Inventory Management & Optimal production levels, purchasing period management, real-time visibility & Reduced excess stock, minimized stockouts, cost savings & Nike (accurate demand forecasting), Ford Motor (inventory management) \cite{13} \\
\hline
Logistics \& Transportation & Route optimization, freight scheduling, real-time tracking, autonomous vehicles & Faster delivery, reduced fuel consumption, lower operational costs, improved efficiency & UPS (ORION system saves 38M liters fuel/yr), Amazon (autonomous robots) \cite{30} \\
\hline
Risk Management & Disruption simulation, anomaly detection, contingency planning, predictive insights & Proactive mitigation, enhanced resilience, improved decision-making & GenAI anticipates risks (congested ports, vendor mapping) and recommends actions \cite{72} \\
\hline
Quality Control & Automated visual inspection, defect detection via sensor data analysis & Reduced recalls, improved product quality, enhanced customer satisfaction & Ford Motor (identifying car seat wrinkles) \cite{66} \\
\hline
Network Design & Optimal configuration identification, scenario evaluation & Efficient network arrangement, balanced cost, availability, and risk & GenAI generates and evaluates various network designs \cite{72} \\
\hline
\end{tabular}}
\end{table}

\subsection{Vulnerabilities and Geopolitical Risks in AI-Integrated Supply Chains}

While Generative AI offers substantial benefits for supply chain optimization and resilience, its integration also introduces new, complex vulnerabilities and exacerbates existing geopolitical risks. The increasing reliance on AI systems creates new points of fragility within global supply chains.

One significant concern is the increased complexity and interdependencies within AI-integrated supply chains. AI systems often rely on a vast ecosystem of software libraries, frameworks, and external Application Programming Interfaces (APIs). If any of these dependencies are compromised or maliciously altered, it can lead to \textit{supply chain attacks,} creating vulnerabilities within the AI system itself and potentially enabling unauthorized access or control by malicious actors \cite{69}.

Data-related risks are also paramount. The quality of GenAI outputs is heavily dependent on the integrity and comprehensiveness of its training data. If this data is incomplete, biased, or flawed, the AI's outcomes will inevitably mirror these issues, leading to skewed or unreliable outputs \cite{14}. Furthermore, the sheer volume of data used for training GenAI models raises significant concerns about sensitive data disclosure or leakage, posing a substantial risk to privacy and security \cite{69}.

Cybersecurity threats are amplified in AI-integrated supply chains. AI systems are susceptible to adversarial attacks, where subtle modifications to input data can deceive AI models, leading to incorrect outputs or decisions \cite{14}. Model poisoning, which involves manipulating the training data itself, can compromise the integrity and performance of AI systems, leading to biased outcomes or service disruptions \cite{14}. Moreover, GenAI can be misused to automate and enhance sophisticated cyber threats, such as generating highly realistic phishing attacks or advanced malware that evades traditional detection \cite{14}. These evolving cyber threats pose significant risks to the overall integrity and functionality of supply chains \cite{71}.

The geopolitical impact of AI chip shortages presents a critical vulnerability. The convergence of surging AI demand, fragile semiconductor manufacturing capacity, and ongoing geopolitical headwinds is placing unprecedented stress on the global supply chain \cite{72}. U.S. export restrictions on advanced AI chips, particularly targeting China, and China's retaliatory measures, such as hoarding critical materials like gallium and germanium, have created a \textit{semiconductor crossroads} \cite{71}. This has resulted in severe advanced node shortages and material constraints, directly impacting global supply chain resilience and industrial production \cite{71}.

Compounding these issues is the geographic concentration of production for critical components. For instance, South Korea's dominance in DRAM production and Taiwan's preeminence in advanced chip fabrication create single points of failure, rendering the global supply chain highly vulnerable to regional disruptions, whether from natural disasters or geopolitical tensions \cite{71}.

The paradox of AI-enhanced resilience and AI-introduced fragility is a central theme in this analysis. While GenAI offers unprecedented tools for improving supply chain resilience through predictive capabilities and optimization, its very integration introduces new, complex points of fragility and geopolitical leverage. The reliance on advanced AI chips, often produced by a limited number of dominant players in geopolitically sensitive regions, transforms the semiconductor supply chain into a \textit{key battleground} in geopolitical rivalry \cite{18}. This creates a situation where the more AI is integrated into supply chains to enhance resilience, the more those supply chains become susceptible to AI-specific vulnerabilities and the geopolitical weaponization of technology, such as export controls and sanctions \cite{32}.

\subsection{Data Sovereignty and its Influence on Cross-Border Data Flows}
Data sovereignty has evolved from a peripheral legal notion into a central strategic priority for AI-driven enterprises, positioning data as a \textit{regulated asset} with profound geopolitical and economic implications \cite{51}. It refers to the principle that data generated or stored within a specific country or jurisdiction must comply with that entity’s legal and regulatory framework \cite{53}.
This growing emphasis on national data control introduces substantial friction into the traditional AI innovation paradigm, which relies on access to large-scale, diverse, and globally distributed datasets \cite{51}. As a consequence, developers may be forced to retrain or segment AI models for different jurisdictions, significantly increasing the cost, complexity, and latency of AI development and deployment \cite{51}. In highly regulated domains—such as finance, healthcare, and defense—strict data localization laws already shape where and how AI models can be trained, fine-tuned, and operationalized \cite{51}. This shift not only complicates the technical pipeline but also challenges the scalability and generalizability of AI systems across borders.

Geopolitical motivations are a primary driver behind the push for data localization:
\begin{itemize}
  \item \textbf{National Security:} Restricting the flow of domestically generated data across borders to geopolitical adversaries is seen as a crucial measure to prevent cyber espionage, the theft of trade secrets, and the launch of offensive cyber operations \cite{54}.

  \item \textbf{Economic Value and \textit{Data Colonization}:} Many developing nations seek to localize data to empower their local companies to ascend the AI value chain. They view the unrestricted flow of data to developed states, which house the majority of the world's data centers and processing capabilities, as a form of \textit{data colonization} \cite{54}. This asymmetry places developing countries at a significant disadvantage in the AI economy if they cannot access and utilize data created within their own borders \cite{54}.

  \item \textbf{Digital Sovereignty:} Countries are increasingly investing in \textit{sovereign AI infrastructure}—cloud-neutral, auditable, and geographically distributed—and developing \textit{sovereign clouds} to ensure control over their data privacy, protection, and storage regulations without external interference \cite{51}. Germany is an example of a country leading this trend \cite{53}.
\end{itemize}

However, data localization also introduces a range of complex challenges. By restricting cross-border data flows, it can hinder economic growth, exacerbate compliance burdens, and potentially empower authoritarian regimes through enhanced surveillance capabilities \cite{54}. Moreover, it contributes to the fragmentation of the global internet, undermining the open data ecosystems upon which much of AI innovation depends. From a technical standpoint, localization demands careful orchestration of where data is generated, collected, processed, and stored are significantly complicating data management practices for AI developers and enterprises alike \cite{53}.

To navigate these challenges, organizations are increasingly adopting technical strategies such as distributed infrastructure and federated AI. Federated approaches involve training smaller models locally at the edge: closer to where data is generated, thereby reducing the need for transferring sensitive data across borders while still extracting valuable insights \cite{53}. This method not only helps meet legal compliance standards but also enhances user privacy and preserves data sovereignty. Another emerging concept is data delocalization, wherein datasets are split and stored across multiple jurisdictions, or even orbital platforms in the case of space-based cloud storage, thereby mitigating risks associated with centralized storage and enhancing cyber resilience \cite{74}.

Nonetheless, the proliferation of such sovereign data strategies is contributing to a broader fragmentation of the global digital economy. As nations assert greater control over domestic data flows, driven by the strategic value of AI and the recognition that data is a critical national asset, the development and deployment of AI systems are being fundamentally restructured. This is not simply a matter of regulatory compliance—it represents a paradigmatic shift that could reshape global technological collaboration. The emergence of \textit{data borders}, denoted by artificial constraints on the free flow of information threatens to hinder the scalability, interoperability, and generalizability of AI models that rely on diverse, cross-regional datasets \cite{51}. Over time, this high volume of data may reduce global efficiency, limit innovation, and disrupt international trade and digital supply chains, posing profound implications for both economic development and technological progress \cite{54}.

\section{Industrial Competitiveness in the GenAI Era}
Generative AI is a powerful catalyst reshaping industrial competitiveness across a multitude of sectors. Its influence extends from optimizing existing processes to accelerating innovation, with significant implications for economic disparities and the strategic importance of talent and infrastructure.

\subsection{Case Studies in Manufacturing, Healthcare, and Logistics}
GenAI is fundamentally reshaping industries by enabling machines to create original content, streamline processes, and enhance innovation across various sectors \cite{62}. Businesses are rapidly adopting this technology to transform traditional models, from personalized marketing campaigns to automated product design \cite{62}.

\subsubsection{Manufacturing}
Generative AI is reshaping the manufacturing sector by enhancing efficiency, precision, and innovation across production processes. From product design to supply chain optimization, GenAI enables data-driven decision-making and real-time responsiveness. These capabilities not only reduce operational costs but also support the transition toward more sustainable and adaptive manufacturing ecosystems. Current use cases are:

\begin{itemize}
  \item \textbf{Product Design \& Development:} GenAI significantly accelerates product design and development by helping engineers generate and test numerous design ideas, thereby reducing costs and time-to-market \cite{62}. A notable example is Autodesk's collaboration with Airbus, where GenAI assists in creating lighter, more fuel-efficient jetliners, such as the partition for the Airbus A320 \cite{63}.

  \item \textbf{Quality Improvement:} GenAI leverages machine learning algorithms to analyze sensor data from manufacturing processes, detecting subtle patterns that indicate potential product flaws early in the production cycle. This proactive approach reduces the risk of costly recalls and ensures the delivery of high-quality goods \cite{60}. Ford Motor, for instance, employs AI to identify defects like wrinkles in car seats, enhancing overall product quality \cite{63}.

  \item \textbf{Predictive Maintenance:} GenAI employs advanced machine learning techniques to analyze sensor data from industrial machinery, identifying patterns that predict potential malfunctions. This enables manufacturers to schedule maintenance proactively, minimizing unplanned downtime and optimizing resource allocation, leading to enhanced equipment performance and extended machinery lifespan \cite{60}.

  \item \textbf{Process Optimization \& Supply Chain Management:} GenAI optimizes inventory levels, accurately forecasts demand, and manages purchasing periods by analyzing current and projected pricing. This improves overall efficiency and reduces waste in manufacturing operations \cite{60}.
\end{itemize}

\subsubsection{Healthcare}
Generative AI is driving transformative change in healthcare by enabling faster diagnostics, personalized treatment planning, and drug discovery. By processing vast medical datasets and generating actionable insights, GenAI enhances clinical decision-making and operational efficiency. Its integration into healthcare workflows supports more precise, data-informed, and patient-centric care delivery. Key use cases are:
\begin{itemize}
  \item \textbf{Personalized Treatment Plans \& Drug Discovery:} GenAI analyzes extensive health datasets to identify patterns for tailored treatment plans, significantly enhancing patient outcomes \cite{75}. It also accelerates drug discovery processes by simulating chemical reactions for research applications \cite{76}. Insilico Medicine achieved drug development success, creating a pulmonary fibrosis candidate in less than 18 months using AI \cite{76}.

  \item \textbf{Medical Imaging \& Diagnostics:} AI imaging tools, powered by GenAI, assist radiologists in detecting anomalies in X-rays, MRIs, and CT scans. This enhances diagnostic accuracy and lowers error rates, leading to speedier and more efficient treatment \cite{76}.

  \item \textbf{Administrative Automation:} GenAI automates repetitive administrative tasks such as medical documentation, note-writing, and appointment arrangement, freeing clinicians to dedicate more time to direct patient care \cite{75}.

  \item \textbf{Fraud Detection:} AI optimizes health insurance technology by streamlining claims processing and detecting fraudulent activities, which results in lower expenses for insurance companies \cite{76}.
\end{itemize}

\subsubsection{Logistics}

In the logistics sector, Generative AI is optimizing route planning, demand forecasting, and supply chain coordination. By synthesizing real-time data from multiple sources, GenAI helps anticipate disruptions, reduce delivery times, and enhance resource utilization. These innovations improve operational resilience and responsiveness in an increasingly complex and globalized logistics environment. Some key uses cases are:
\begin{itemize}
  \item \textbf{Operational Efficiencies:} GenAI automates repetitive cognitive tasks, such as billing, payroll, and data entry, and enhances traditional logistics processes, including demand forecasting and supplier negotiations \cite{77}. This allows logistics professionals to focus on higher-value tasks.

  \item \textbf{Route Optimization \& Real-time Scheduling:} AI algorithms process vast amounts of real-time data from logistics networks to select optimal routes, forecast delays, and enhance supply chain visibility \cite{64}. UPS's ORION route optimization system is a prime example, processing thousands of route optimizations per minute \cite{61}.

  \item \textbf{Warehouse Automation:} AI-powered robots and automated systems efficiently handle tasks like picking, sorting, and packaging in distribution centers, reducing reliance on human workers and enabling round-the-clock operations \cite{64}. Amazon extensively utilizes autonomous robots in its warehouses for order fulfillment and inventory tracking \cite{66}.
\end{itemize}

GenAI is speeding up how quickly innovation happens across industries, from manufacturing to healthcare to logistics. It allows companies to rapidly test and improve designs, discover new drugs faster, and streamline their operations with better predictions and automation \cite{62}. As a result, businesses that adopt GenAI early are gaining a major head start. They can bring products to market more quickly, reduce costs, and respond faster to changing demands. This growing lead makes it harder for competitors, especially those slow to adopt GenAI to catch up \cite{63}.
For countries, the stakes are even higher. Nations that invest early in GenAI technologies, support skilled talent development, and build strong digital infrastructure are better positioned to lead in the next wave of industrial and economic growth. These countries are likely to see increased productivity, attract foreign investment, and develop more resilient supply chains. On the other hand, countries that lag behind may face serious challenges—such as losing market share, becoming dependent on foreign AI technologies, or falling further behind in global competitiveness \cite{36}. This widening gap reinforces the urgency for strategic national investments and policies that promote responsible, widespread adoption of GenAI across sectors.

\subsection{National Adoption Rates and Disparities}
Global race to adopt Generative AI is reshaping economic and industrial dynamics, but progress is uneven across regions. While some nations are rapidly integrating GenAI into their economies, others face structural barriers that risk widening existing inequalities. Here, we explore these disparities and their broader implications: 
\subsubsection{Global Patterns of GenAI Adoption}
Global rollout of AI technologies is progressing at a remarkable pace, with 56\% of companies worldwide reporting adoption of AI in at least one business function \cite{36}. However, the picture becomes more nuanced when we examine Generative AI (GenAI) adoption specifically. North America currently leads GenAI uptake, with nearly 40\% of surveyed companies reporting use of GenAI tools. Asia-Pacific follows closely, benefiting from strong investment in digital transformation and innovation ecosystems, boasting 55\% adoption of AI more broadly \cite{36}. Europe reports a 48\% adoption rate, driven by coordinated public-private initiatives, while the Middle East and Latin America lag behind at 35\% and 28\%, respectively \cite{36}.

These regional disparities are shaped by multiple structural factors. Countries with robust digital infrastructure, such as widespread cloud services, broadband connectivity, and access to high-performance computing, are better positioned to scale GenAI deployment. The presence of a digitally skilled workforce, particularly in machine learning and data engineering, is another critical enabler \cite{36}. In contrast, countries with limited investment in digital infrastructure or lacking in skilled talent face slower AI diffusion. Furthermore, regulatory clarity plays a decisive role—nations with adaptive but predictable AI governance environments tend to attract greater experimentation and private-sector investment.
While the global AI adoption trend is upward, its uneven distribution raises concerns about the consolidation of technological advantage within certain regions. Without coordinated international efforts to support equitable access to GenAI tools and training, the current trajectory could deepen global technological divides.

\subsubsection{Structural Inequality and the Digital Divide}
AI adoption is not unfolding on a level playing field. High-income countries have a head start due to several pre-existing advantages: mature innovation ecosystems, deep pools of investment capital, and extensive digital infrastructure \cite{48}. These countries also benefit from access to large, high-quality datasets and world-class research institutions. Consequently, they continue to attract the lion’s share of global AI-related investment, reinforcing their lead. This dynamic draws resources away from developing nations, many of which lack the infrastructure or human capital to compete at scale \cite{47}.
The result is a growing structural imbalance where wealthier nations realize disproportionate gains from AI integration, while developing nations risk becoming consumers rather than producers of AI technology. This phenomenon can lead to what some analysts describe as \textit{technological dependency,} with lower-income countries becoming reliant on AI tools and platforms developed and owned by foreign entities \cite{47}. In macroeconomic terms, this may manifest as a transitional decline in GDP for developing countries due to reduced competitiveness in high-tech and services sectors \cite{47}.
Moreover, the digital divide exacerbates vulnerability. Nations without strong data protection frameworks or sovereign digital infrastructure may find themselves at a disadvantage in shaping how GenAI systems are applied or governed. To close this gap, targeted investments in connectivity, education, and data sovereignty are essential.

\subsubsection{Productivity Gains vs. Labor Market Disruption}
While GenAI promises large productivity gains for businesses and economies, it also raises serious questions about its impact on employment and income distribution. GenAI technologies are particularly well-suited to automating cognitive and semi-cognitive tasks, which expands the range of jobs susceptible to displacement. For example, research suggests that more than 90\% of tasks performed by logistics managers could be automated using GenAI tools \cite{77}. While such automation can reduce costs and improve efficiency, it may simultaneously displace middle-income workers, contributing to wage stagnation or job loss \cite{77}.
This disruption could deepen socioeconomic inequality. Workers with advanced technical skills and digital literacy will benefit most, while those in routine or cognitively repetitive roles may find their skills devalued. This uneven distribution of AI benefits reflects a broader trend—the so-called \textit{productivity-inequality trade-off} \cite{62}. GenAI enables firms to achieve significant gains in output and efficiency, but those gains are often captured by a small number of firms or individuals who control the underlying infrastructure or intellectual property.
Without proactive labor policies—such as retraining programs, income support, or inclusive innovation strategies—GenAI could reinforce existing power asymmetries in both developed and developing economies. As AI becomes more deeply embedded in the global economy, the challenge will be to ensure that its benefits are shared broadly, and that economic transformation does not come at the expense of social cohesion \cite{48}.

\subsection{Critical Role of AI Talent and Infrastructure in National Competitiveness}
National competitiveness in AI increasingly depends on a country’s ability to develop and sustain both cutting-edge talent and robust infrastructure. These elements are critical for driving innovation, scaling AI technologies, and maintaining strategic autonomy in a rapidly evolving global landscape.

\subsubsection{Strategic Importance of AI Talent}
A skilled workforce is a cornerstone of national AI competitiveness. Nations that excel in cultivating, attracting, and retaining AI talent have a distinct advantage in developing cutting-edge applications and research \cite{25}. This includes not only technical roles such as machine learning engineers and data scientists, but also interdisciplinary experts who can integrate AI into domains like healthcare, logistics, and public policy. Countries like South Korea are proactively investing in this area by doubling AI-related undergraduate and graduate seats and revamping their curriculum to emphasize practical and applied AI knowledge \cite{42}. Additionally, national strategies must extend beyond elite education to include vocational training and continuous upskilling programs to prepare broader segments of the workforce for AI integration \cite{25}. Failure to address the talent gap can limit a nation’s ability to innovate and scale AI solutions. In contrast, nations that prioritize AI education can position themselves as global leaders in both AI development and adoption.

\subsubsection{Foundational Infrastructure and Compute Power}
Infrastructure forms the backbone of AI capabilities. Training and deploying AI models—especially large-scale GenAI systems—require advanced semiconductors, powerful computing clusters, extensive energy resources, and high-speed digital networks \cite{21}. Data centers, in particular, have become strategic national assets, leading governments to promote local investments and cloud-region development \cite{80}. The energy demands of AI training are substantial, driving parallel efforts in building sustainable power solutions and optimizing computational efficiency \cite{21}. The U.S., for instance, has invested heavily in high-performance computing hubs and is partnering with industry to scale AI-ready infrastructure. However, infrastructure gaps remain stark between high-income and low-income countries, deepening technological inequalities. Without sufficient infrastructure, even nations with capable talent pools may struggle to develop or deploy AI at scale \cite{49}.

\subsubsection{Emerging Infrastructure Arms Race}

As AI becomes a driver of economic and strategic influence, a global infrastructure arms race is emerging. Countries are vying for dominance not only in human capital but also in physical assets like semiconductor fabrication plants, GPU supply chains, and hyperscale data centers \cite{49}. This race is compounded by geopolitical conflicts: for example, the U.S.–China tech rivalry has escalated export controls on advanced chips and AI components, disrupting global supply chains and raising barriers to entry \cite{49}. Such tensions have also driven up construction costs for U.S. data centers and incentivized reshoring of critical manufacturing \cite{49}. Strategic autonomy in AI now requires full-stack capabilities—from chip design to cloud deployment—otherwise, nations risk becoming dependent on foreign platforms and vulnerable to supply chain disruptions \cite{21}. As this race intensifies, disparities between nations with AI sovereignty and those without may widen, reshaping the global technology order.

\section{Evolving Governance Models}
Rapid advancement and widespread diffusion of Generative AI (GenAI) and autonomous systems have made AI governance a critical global imperative. This section explores the diverse and often fragmented approaches to AI governance, examining national frameworks, international efforts, the impact of export controls, and the ongoing pursuit of ethical guidelines.

\begin{table*}[t]
\caption{Overview of International AI Governance Frameworks}
\label{tab:ai-governance-frameworks}
\scalebox{0.8}{
\begin{tabular}{|p{2cm}|p{2cm}|p{3cm}|p{4.5cm}|p{4.5cm}|}
\hline
\textbf{Framework/ Initiative} & \textbf{Type} & \textbf{Primary Focus/ Scope} & \textbf{Key Features/ Mechanisms} & \textbf{Challenges/Limitations} \\
\hline
EU AI Act & Binding Legislation & Human-centric, ethical AI, risk mitigation, market access & Risk categorization (unacceptable, high, limited, minimal), strict requirements for high-risk systems, transparency, human oversight, regulatory sandboxes & Geopolitical tension with other blocs, potential for market entry barriers, implementation complexity across member states  \cite{14} \\
\hline
OECD Principles on AI & Non-binding Principles & Transparent, explainable, and accountable AI development & Principles for responsible AI, framework for AI governance, broad adherence by member and non-member countries & Non-binding nature, limited enforcement power, geopolitical tensions can impede global frameworks  \cite{54} \\
\hline
UN HLAB (High-level Advisory Body on AI) & Advisory Body/Report & Global AI governance for humanity, equitable distribution of benefits/risks & Fostering consensus, multidisciplinary approach, AI Standards Exchange, addressing inequalities & Under-cut by political division, exclusion of major AI stakeholders (e.g., China, Russia), risk of ``AI colonialism''  \cite{41} \\
\hline
ICRC AI Policy & Policy Guidance & Responsible use of AI in armed conflict, IHL compliance & Meaningful human control in military decisions, adherence to international humanitarian law, ``do no harm'' principle & Geopolitical tensions delay global consensus on military AI, lack of public awareness on risks  \cite{58} \\
\hline
G7 Hiroshima AI Process & Voluntary Commitments & Safe, secure, and trustworthy AI & Voluntary code of conduct, international cooperation & Non-binding, limited scope, geopolitical tensions can influence and impede efforts  \cite{54} \\
\hline
Council of Europe’s Framework Convention on AI & Inter-governmental Agreement & Human rights, democracy, rule of law in AI & Legal framework for AI governance & Impact and enforcement still evolving, potential for fragmentation with other frameworks  \cite{4} \\
\hline
\end{tabular}}
\end{table*}

\subsection{National Governance Frameworks and Regulatory Challenges}
AI governance encompasses the frameworks, policies, and practices designed to promote the responsible, ethical, and safe development and use of AI systems \cite{82}, aiming to balance public safety and civil rights with the need to foster innovation \cite{83}. Different nations have adopted varied approaches reflecting their political, economic, and cultural contexts. The European Union (EU) has implemented the EU AI Act, the world’s first comprehensive, binding, and cross-sectoral legal framework for AI \cite{33}, which emphasizes transparency, risk mitigation, and regulatory oversight. It categorizes AI systems into four risk levels—unacceptable, high, limited, and minimal—with high-risk systems, such as those used in biometrics or critical infrastructure, required to meet stringent conditions including risk assessments, human oversight, and thorough technical documentation \cite{39}. The EU’s primary goal is to promote human-centric and trustworthy AI \cite{16}. In contrast, the United States (U.S.) adopts a more fragmented and reactive approach, focusing on sector-specific regulations rather than a unified legal framework \cite{39}. The U.S. emphasizes fostering innovation and leveraging AI leadership to bolster national security, with federal initiatives prioritizing long-term R\&D investment, workforce development, technical standards, and building public trust in AI \cite{35, 21}. China’s strategy is distinctly state-led, integrating AI into social governance with a proclaimed human-centric regulatory stance and specific rules for Generative AI and deep synthesis technologies alongside broader policy frameworks \cite{56, 16}. Meanwhile, Japan prefers a soft-law approach that promotes voluntary, nonbinding business initiatives and guidelines rather than strict legal regulations \cite{41}. These diverse governance models highlight differing national priorities and strategies in managing the complex opportunities and risks presented by AI technologies. Table \ref{tab:ai-governance-frameworks} summarizes key national and international AI governance frameworks, highlighting their distinct regulatory approaches and strategic priorities.

Despite these varied approaches, common challenges persist across national governance frameworks:
\begin{itemize}
  \item \textbf{Lack of Transparency (Black Box Problem):} Many GenAI models operate as \textit{black boxes,} making it difficult to discern how they arrive at their decisions or outputs. This opacity hinders accountability and makes it challenging to build trust, particularly in critical applications where understanding the reasoning process is essential \cite{5}.

  \item \textbf{Bias and Discrimination:} GenAI models can inadvertently learn and amplify biases present in their training data, leading to discriminatory outcomes in areas like language generation or decision-making systems \cite{14}.

  \item \textbf{Data Privacy and Security:} The reliance of GenAI on vast amounts of data raises significant concerns about sensitive data leakage, misuse, and vulnerability to cyberattacks, necessitating robust protection measures \cite{14}.

  \item \textbf{Hallucinations and Inaccuracies:} GenAI systems can produce fabricated or incorrect results, a phenomenon known as \textit{hallucination,} which is highly problematic in high-risk applications such as healthcare, defense, or the judiciary \cite{14}.

  \item \textbf{Jailbreaking:} This refers to the manipulation of GenAI models to bypass their built-in guardrails, eliciting responses that would normally be refused \cite{31}.

  \item \textbf{Regulatory Lag:} The rapid evolution of AI technology frequently outpaces existing legal frameworks, making it difficult for regulations to keep pace with technological advancements \cite{5}.
\end{itemize}

The \textit{Brussels Effect} versus the \textit{Silicon Valley Effect} represents two competing models of global AI governance, creating a notable \textit{regulatory tension} \cite{39}. The EU, through its comprehensive and binding AI Act, aims to project normative power and set international benchmarks, influencing global standards through its large single market \cite{33}. This is often referred to as the \textit{Brussels Effect.} In contrast, the U.S. favors a more deregulated, market-driven approach, aiming to spread its standards through technological innovation, market dominance, and voluntary industry guidelines—the \textit{Silicon Valley Effect} \cite{33}. The effectiveness of these differing approaches in shaping global AI development and deployment, and their geopolitical implications, is a key area of contestation. This divergence forces companies operating across jurisdictions to navigate a fragmented compliance environment, potentially leading to \textit{regulatory arbitrage} where firms choose to operate in regions with more lenient AI laws \cite{39}.

\subsection{International Regulatory Efforts and the Formation of AI Blocs}
Achieving global consensus on AI regulations and frameworks faces significant challenges, primarily due to prevailing geopolitical tensions \cite{16}. The absence of clear, universally agreed-upon definitions for AI technologies and a unifying catalyst, unlike the circumstances that led to international treaties in the nuclear era, complicates the drafting and enforcement of comprehensive international agreements \cite{45}. 

Despite these hurdles, several international initiatives are underway:
\begin{itemize}
    
\item \textbf{OECD Principles on AI:} These principles provide a framework for the transparent, explainable, and accountable development and deployment of AI \cite{87}. Many non-member countries have adhered to these principles, indicating a degree of international alignment \cite{21}.

\item \textbf{UN Secretary-General's High-level Advisory Body on AI (HLAB):} The HLAB advocates for a robust global framework for AI governance, acknowledging the unequal distribution of AI's benefits and risks worldwide \cite{27}. It proposes fostering consensus, adopting a multidisciplinary approach, and harmonizing AI standards through an \textit{AI Standards Exchange} \cite{27}. However, the HLAB's efficacy is hindered by political divisions and the exclusion of major AI stakeholders, such as China and Russia, from key discussions \cite{45}.

\item \textbf{ICRC AI Policy:} The International Committee of the Red Cross (ICRC) has developed an AI Policy to guide the responsible use of AI, particularly emphasizing the critical need for \textit{meaningful human control} in military decisions and adherence to international humanitarian law (IHL) \cite{16}.

\item \textbf{G7 Hiroshima AI Process:} This initiative introduced voluntary commitments aimed at ensuring \textit{safe, secure and trustworthy AI} \cite{89}.

\item \textbf{Council of Europe's Framework Convention on Artificial Intelligence:} This represents another intergovernmental agreement contributing to the evolving governance landscape \cite{21}.
\end{itemize}

Global race for AI supremacy is catalyzing the emergence of distinct \textit{innovation blocs} and strategic alliances, as states align based on technological capabilities, regulatory philosophies, and geopolitical interests \cite{22}. This intensifying competition carries risks: in the pursuit of dominance, some nations may cut corners on safety and ethical safeguards, favoring rapid deployment over responsible governance \cite{49}. Within this landscape, the private sector wields disproportionate influence. Major technology firms, as primary engines of AI innovation and massive consumers of computational resources, play a dual role—driving progress while also shaping the governance environment to align with corporate interests \cite{74}. Their substantial lobbying power can stall or weaken regulatory efforts, with many preferring voluntary ethical frameworks that offer flexibility without legal enforcement \cite{46}. In contrast, civil society organizations (CSOs) are critical counterweights in the governance discourse. These actors—ranging from advocacy groups to academic and policy think tanks—amplify voices from marginalized communities, promote transparency and algorithmic fairness, and hold both governments and corporations accountable \cite{56}. CSOs also contribute to public literacy on AI risks and benefits and participate in international arms control dialogues, particularly around the militarization of AI, through rigorous research and monitoring \cite{91}. However, despite these efforts, the global governance of AI remains patchy and underdeveloped. The ongoing failure to establish comprehensive, binding international frameworks has given rise to \textit{minilateralism}, where small coalitions of like-minded countries attempt to set shared norms and technical standards \cite{21}. Although such efforts can serve as valuable incubators for policy experimentation and consensus-building, they risk deepening global fragmentation by creating competing regulatory regimes with limited interoperability \cite{45}. This fragmentation not only hampers global coordination on safety, ethics, and trade but also intensifies geopolitical divides. Countries excluded from dominant alliances may forge parallel AI ecosystems, which could reinforce techno-nationalist agendas and diminish opportunities for global cooperation on existential challenges posed by advanced AI systems \cite{55}.

\subsection{Export Controls, Technology Transfer, and Geopolitical Outcomes}
As AI becomes a strategic asset, nations are leveraging export controls and technology transfer policies to secure competitive advantage and mitigate security risks \cite{30}. These measures are shaping not only the pace and direction of AI development, but also the global balance of power. In this section, we examine the use of export controls as tools of national security, China’s strategic response, the emergence of circumvention tactics, and the broader geopolitical consequences of supply chain weaponization.

\subsection{Export Controls, Technology Transfer, and Geopolitical Outcomes}

\subsubsection{Export Controls as National Security Tools}
Artificial intelligence is now central to national security and economic competitiveness, with the most powerful systems posing significant risks if accessed by adversaries \cite{30}. To mitigate such risks, nations—particularly the United States—are increasingly turning to export controls on AI-related technologies as geopolitical instruments. These include restrictions on the export of advanced AI chips, semiconductor manufacturing tools, and specialized software to countries like China \cite{30, 18}. The goal is to curtail the ability of targeted nations to develop and deploy advanced AI models that could threaten global stability or shift the balance of technological power. Such measures are embedded within broader strategies to maintain leadership in critical technologies while influencing global supply chains. Export controls thus serve a dual purpose: blocking access to sensitive technology and reinforcing alliances through selective technology sharing \cite{96, 97}. However, these controls also raise concerns about market distortion and long-term global collaboration on AI safety.

\subsubsection{China’s Strategic Response and Innovation Push}
While U.S. export controls have constrained China’s access to high-end chips and manufacturing equipment, they have also galvanized a significant domestic innovation drive \cite{93}. Major Chinese firms like SMIC, Huawei, and CXMT are investing in indigenous chip design and production capabilities to reduce foreign dependency \cite{95}. Additionally, Chinese developers are turning to open-source Large Language Models (LLMs) like DeepSeek to sidestep hardware limitations and continue GenAI development at scale \cite{17}. This push for self-reliance is supported by significant state funding and national AI development strategies. As a result, rather than deterring advancement, export controls may be catalyzing China’s long-term resilience and its ambition for technological independence. While the short-term impact includes reduced performance and supply bottlenecks, the long-term implications could involve the rise of a parallel AI superpower with a separate ecosystem.

\subsubsection{Circumvention, Loopholes, and Global Spillovers}
Despite efforts to enforce strict export restrictions, several workarounds have emerged that undermine their effectiveness. Chinese entities reportedly use cloud computing services hosted abroad to access restricted U.S. AI chips, allowing them to train models without direct hardware acquisition \cite{32}. In addition, shell companies are allegedly being used to build data centers in third countries that are not under the same export restrictions, enabling indirect access to restricted technologies \cite{94}. These loopholes highlight the difficulty of enforcing unilateral controls in a globally interconnected technology landscape. Moreover, these restrictions often impose unintended burdens on U.S. universities, increasing compliance costs and complicating international research collaboration \cite{32}. Over-regulation may also shift AI innovation further toward the private sector, diminishing the public sector’s role in foundational research and limiting the broader benefits of academic inquiry.

\subsubsection{Weaponization of Supply Chains and Multipolar Outcomes}
Export controls have accelerated the weaponization of global supply chains, turning semiconductors and advanced computing resources into tools of strategic leverage \cite{18}. The U.S. increasingly uses these controls not only to restrict adversaries but also to reward allies, fostering global alignment with its standards and ecosystem \cite{96, 97}. However, this strategy also intensifies the race for technological independence, as countries like China respond by bolstering domestic infrastructure, R\&D, and AI talent pipelines \cite{94}. The resulting fragmentation contributes to the emergence of competing AI ecosystems with diverging technical standards, regulatory philosophies, and ethical norms. This shift toward a multipolar AI world raises the risk of reduced global coordination on safety, interoperability, and governance. Ultimately, while export controls serve immediate geopolitical objectives, they may catalyze longer-term divergence that complicates efforts to ensure AI remains a shared and beneficial technology.

\subsection{Ethical Guidelines and the Pursuit of Human-Centric AI}
As Generative AI and autonomous systems continue to advance, ensuring they align with fundamental human values is more urgent than ever. Ethical governance of AI focuses on fairness, transparency, accountability, and privacy, all crucial to fostering public trust and safeguarding against unintended harms \cite{82,98}. This section explores six key facets of human-centric AI ethics: foundational principles, implementation challenges, mitigation strategies, privacy protection, geopolitical dimensions, and global values competition.

\subsubsection{Foundational Ethical Principles}
Most AI ethical frameworks converge on a core set of values that include fairness and bias mitigation, transparency and explainability, accountability, and privacy and security \cite{82}. Fairness requires that AI systems do not produce discriminatory outcomes, particularly in sensitive domains like hiring, lending, or law enforcement \cite{85}. Bias in AI models often originates from skewed or incomplete training data, necessitating the use of diverse and representative datasets \cite{85}. Transparency is another central concern—AI decisions must be interpretable by humans, especially for systems deployed in high-stakes environments \cite{82}. This entails clear documentation of model behavior and decision rationale. Accountability mechanisms are vital, with explicit assignment of responsibility and human oversight for critical applications \cite{82}. Lastly, strong protections must be in place for data privacy, including consent protocols and cybersecurity safeguards \cite{85}. Together, these principles form the ethical bedrock of trustworthy AI.

\subsubsection{Challenges in Implementation}
Despite wide recognition of these ethical imperatives, operationalizing them is fraught with complexity. One of the most persistent issues is the difficulty of creating universal ethical standards that are culturally and geopolitically neutral \cite{86}. AI systems developed in one region may embed assumptions or norms that are not shared elsewhere. Another challenge lies in the \textit{black box} nature of many AI models—particularly deep learning systems—where decisions cannot easily be explained or audited \cite{15}. This opacity undermines accountability and erodes user trust. Furthermore, the pace of AI innovation frequently outstrips the evolution of ethical frameworks and enforcement mechanisms \cite{85}. Regulatory bodies and institutions often struggle to keep up with the risks posed by fast-developing GenAI capabilities. These implementation barriers highlight the gap between ethical theory and practice in the AI domain.

\subsubsection{Practical Mitigation Strategies}

To bridge the implementation gap, a number of proactive mitigation strategies are being explored. First, bias audits should be institutionalized throughout the AI development lifecycle, not just post-deployment \cite{82}. Fairness must be assessed at multiple levels: data, model, and decision outcome. Second, explainability tools—like saliency maps or decision trees—can help open the black box and make AI decisions more comprehensible \cite{82}. For high-risk applications, a principle of \textit{meaningful human control} should be enforced, clearly defining when and how humans intervene in automated decisions \cite{11}. Data protection protocols should incorporate state-of-the-art techniques like homomorphic encryption, differential privacy, and access-limiting architectures \cite{85}. Finally, ethical AI cannot be built in silos—interdisciplinary collaboration among technologists, ethicists, sociologists, and legal scholars is essential \cite{98}. These strategies are not panaceas but offer pragmatic pathways to more ethically grounded AI systems.

\subsubsection{Data Privacy and Consent Mechanisms}
Privacy is one of the most deeply contested ethical terrains in GenAI development. These systems require massive volumes of data to function effectively, often including sensitive personal or behavioral data \cite{85}. Without rigorous controls, this creates vulnerabilities related to surveillance, misuse, and data breaches. Ethical AI frameworks advocate for strong consent mechanisms, ensuring users understand and agree to how their data is used \cite{82}. However, obtaining truly informed consent is difficult in complex, opaque systems. Technical solutions like data minimization, anonymization, and encryption must be paired with legal frameworks that enforce user rights \cite{85}. Additionally, transparency in data sourcing and the right to opt out of AI-driven profiling are essential components of privacy-respecting AI. In sum, privacy is not just a technical or legal concern but a human rights issue in the age of ubiquitous algorithmic systems.

\subsubsection{Geopolitical Dimensions of Ethics}

Ethical AI is not merely a question of principles—it is deeply entangled with geopolitical power dynamics. Some governments may downplay ethical concerns in favor of economic gains or political control, especially in authoritarian regimes where civil liberties are already constrained \cite{21}. Although both the European Union and China promote a \textit{human-centric} approach to AI, the philosophical underpinnings of their strategies differ significantly \cite{16}. The EU emphasizes individual rights and democratic oversight, while China’s model places greater emphasis on social harmony and state-led governance. This divergence risks turning ethical frameworks into geopolitical instruments, used to assert normative dominance on the global stage. For example, the UN’s promotion of \textit{trustworthy} AI may seem neutral, but could reflect Western priorities and marginalize non-Western conceptions of ethics \cite{55}. In this way, AI ethics becomes both a mirror and a battleground for global political values.

\subsubsection{Ethics as a Global Values Competition}
Ultimately, the governance of AI is a proxy for a deeper struggle over which values will shape the future of human-machine interaction. Nations are using AI policies to implicitly project their ideologies—liberal democracies prioritize freedom and privacy, while others prioritize control and efficiency \cite{99}. This competition over values informs ethical norms related to autonomy, accountability, and transparency. It also affects the kinds of technologies that are developed and how they are deployed. As different regions codify their own ethical principles, we risk the emergence of fragmented AI ecosystems that reflect distinct political ideologies \cite{55}. This fracturing could hinder cross-border collaboration and make it harder to develop unified responses to global challenges like disinformation, AI misuse, or surveillance capitalism. Therefore, ethical AI is not just a technical challenge but a civilization-level negotiation over the soul of our digital future.

\section{Discussion}
The analysis presented in this paper underscores that Generative AI (GenAI) and autonomous systems are not merely technological advancements but potent geopolitical factors fundamentally reshaping Industry 5.0. These capabilities are rapidly becoming strategic national assets, influencing global supply chain resilience, industrial competitiveness, and demanding the evolution of new governance models. The discussion below integrates the preceding themes—ethical, geopolitical, industrial, and governance-related—into a coherent framework, highlighting the risks and imperatives shaping the future of AI in a multipolar world.

\subsection{Human-Centric Paradox in Industry 5.0}

Industry 5.0’s ideal of human-centric automation increasingly contends with the rise of autonomous GenAI systems that may supersede, rather than support, human agency \cite{11,82}. While human-machine collaboration is a foundational premise, meaningful human control is eroded when decision-making shifts from augmentation to delegation. GenAI systems, particularly those used in logistics, finance, and healthcare, are already executing decisions with limited human oversight, raising profound ethical and practical questions \cite{85,98}. The growing opacity of AI models—the so-called \textit{black box} problem—complicates efforts to ensure transparency, accountability, and auditability \cite{15}. As these systems evolve, policy frameworks must adapt dynamically to embed explainability, fairness, and redress mechanisms into both design and deployment \cite{82}. The challenge is especially acute in high-stakes applications, where lapses in AI accountability may translate into real-world harm. The ethics of GenAI are not merely theoretical constructs but governance imperatives with direct consequences for social cohesion and public trust. Crucially, ethical guidelines must reflect pluralistic values and cultural contexts to avoid the imposition of hegemonic standards under the guise of universalism \cite{86}. Without this adaptability, ethical AI may become exclusionary or politically charged, undermining its legitimacy and effectiveness. Thus, the pursuit of human-centric AI demands not only technical solutions but also continuous political and societal negotiation \cite{98}.

\subsection{AI as a Geopolitical Asset and Technological Divider}

The global proliferation of GenAI has inaugurated a new phase of techno-political competition, where AI is both a tool of statecraft and an arena for ideological contestation \cite{21,55}. The so-called \textit{AI arms race} now extends well beyond military contexts, encompassing economic influence, regulatory soft power, and sociopolitical control \cite{99}. Leading AI nations are establishing distinct technological spheres, characterized by incompatible regulatory standards, localized data infrastructure, and divergent ethical principles \cite{16}. Export controls on advanced semiconductors, AI chips, and foundational models—particularly by the United States—are not just commercial instruments but geopolitical levers intended to constrain adversarial capabilities and preserve strategic dominance \cite{21}. These restrictions have prompted targeted nations, such as China and Russia, to double down on technological self-reliance and domestic innovation ecosystems, accelerating the formation of parallel AI ecosystems. The resulting bifurcation reflects a deeper \textit{values competition}, wherein regulatory choices become proxies for differing visions of autonomy, freedom, and control \cite{99}. This competition also manifests in what some scholars describe as \textit{AI colonialism}, wherein dominant tech powers export their architectures, standards, and ideologies to the Global South, often marginalizing indigenous knowledge systems or governance traditions \cite{55}. Consequently, AI is no longer a neutral tool—it is a geopolitical actor embedded with the intentions and ideologies of its creators. The emerging multipolar AI world challenges the possibility of a unified technological future and intensifies the urgency of global coordination.

\subsection{Double-Edged Sword of AI-Enabled Supply Chains}

GenAI’s transformative impact on global supply chains offers both unprecedented optimization potential and a set of new systemic vulnerabilities. AI can convert traditional reactive supply chain models into anticipatory, self-correcting systems by leveraging real-time analytics, demand forecasting, and autonomous routing \cite{98}. However, this interconnectivity introduces complex interdependencies that can become critical failure points in the event of disruptions, data corruption, or cyber-attacks \cite{85}. The increased reliance on centralized data hubs and proprietary AI platforms heightens the risks of monopolistic control and single points of failure. Furthermore, as AI systems become integral to supply chain management, they can be weaponized through trade restrictions or export controls on foundational technologies such as chips, cloud infrastructure, and large-scale models \cite{21}. For instance, limiting access to advanced GPUs or model weights forces nations to develop indigenous AI capabilities, leading to strategic decoupling and technological fragmentation. As data sovereignty becomes a strategic priority, cross-border data flows face increasing restrictions, impeding collaborative innovation and inflating compliance costs \cite{55}. This fracturing not only impairs global supply chain resilience but also deepens the divide between digitally mature and emerging economies. Paradoxically, while AI enhances local resilience, it reduces systemic robustness due to a lack of interoperability and redundancy across ecosystems. Policymakers must therefore strike a delicate balance between sovereignty and openness, innovation and regulation.

\subsection{Industrial Competitiveness and the Productivity-Inequality Trade-Off}

GenAI is dramatically accelerating industrial innovation cycles, offering productivity gains through automation, intelligent design tools, and synthetic data generation \cite{82}. Early adopters—particularly in sectors such as pharmaceuticals, finance, and advanced manufacturing—are achieving exponential efficiency and scalability advantages. However, this acceleration comes with a \textit{productivity-inequality trade-off}: regions and firms unable to access, afford, or integrate GenAI tools risk falling behind at an increasing rate \cite{85}. Wealthier nations are reinforcing their industrial leads via AI-driven R\&D, while lower-income countries face widening capability gaps and dependence on external technologies. Within countries, the automation of white-collar jobs is displacing mid-skill labor, exacerbating inequality and intensifying social polarization \cite{98}. In many cases, these displacements disproportionately affect marginalized communities, compounding historical inequities. Moreover, AI-driven consolidation in tech sectors threatens market competition, allowing dominant firms to accumulate disproportionate economic and political power. The industrial benefits of GenAI must therefore be weighed against their socio-economic costs. Policies such as inclusive innovation funds, AI literacy programs, and redistributive taxation could help mitigate these disparities, but they require strong political will and coordinated global action. Without such measures, GenAI may entrench rather than alleviate inequality.

\subsection{Fragmented Governance and the Rise of AI Minilateralism}

In the absence of robust global institutions capable of enforcing cohesive AI norms, the world is witnessing the rise of \textit{minilateralism}—small, strategically aligned coalitions attempting to govern specific aspects of GenAI \cite{99}. Initiatives like the EU’s AI Act, the OECD Principles, and the US-led AI Safety Summits illustrate this trend toward decentralized norm-setting. While these frameworks offer practical responses to urgent ethical and safety concerns, they risk reinforcing geopolitical blocs and deepening regulatory fragmentation \cite{16}. The divergence in values—privacy versus surveillance, individual rights versus state control—makes consensus difficult and weakens the efficacy of international coordination. Moreover, Global South nations often remain excluded from these minilateral arrangements, limiting the legitimacy and representativeness of emerging governance norms \cite{55}. The risk is the entrenchment of a \textit{two-speed AI governance regime}, where advanced nations regulate proactively while others are left with minimal oversight or must import foreign standards. This fragmentation complicates cross-border collaboration on shared challenges such as misinformation, cybercrime, and AI misuse. Ethical convergence, therefore, must not merely be a technocratic goal but a diplomatic imperative requiring inclusive dialogue, capacity building, and institutional innovation. Failing to bridge this governance gap could entrench a fractured, unstable global AI landscape with cascading economic and political consequences.

\section{Policy Recommendations} 

In light of the rapidly evolving role of Generative AI (GenAI) and autonomous systems in shaping industrial competitiveness, global governance, and geopolitical dynamics, coordinated and forward-thinking policy action is critical. As highlighted in the discussion, the GenAI revolution introduces both transformative opportunities and profound risks—from supply chain restructuring and talent concentration to ethical divergence and AI-enabled strategic fragmentation. To ensure that AI serves the broader goals of resilience, inclusion, and human-centric progress, multi-level policy responses are required across national, industrial, and international domains.

\subsection{For Governments}  
National governments must assume a proactive and strategic role in both enabling and regulating AI development. This includes investing in AI infrastructure, safeguarding against techno-economic dependencies, aligning regulations with democratic values, and fostering talent pipelines to close capability gaps. A coherent, adaptive policy strategy is essential to balance innovation with security and ethical responsibility.

\begin{enumerate}
    \item \textbf{Prioritize Human-AI Collaboration Frameworks:} Develop national strategies that explicitly focus on augmenting human capabilities through AI, rather than solely on full automation. This involves investing in education and training programs that foster human-AI teaming, creativity, and critical thinking to ensure a skilled workforce capable of leveraging GenAI effectively in Industry 5.0. \cite{6}

    \item \textbf{Invest Strategically in Sovereign AI Infrastructure and Talent:} Recognize the ``infrastructure arms race'' and ``talent war'' as critical dimensions of national security. Governments should significantly increase public and private investment in domestic AI research, advanced computing infrastructure (data centers, supercomputers), and talent development pipelines to reduce technological dependence and foster homegrown AI capabilities. \cite{25}

    \item \textbf{Develop Adaptive and Nuanced Regulatory Frameworks:} Implement AI governance policies that are flexible enough to keep pace with rapid technological advancements while ensuring ethical deployment. This includes clear guidelines for human oversight in autonomous systems, mechanisms for bias detection and mitigation, and robust data privacy and security protocols. Avoid overly broad export controls that may stifle domestic innovation. \cite{32}

    \item \textbf{Strengthen Supply Chain Resilience through Diversification and AI:} Mandate and incentivize diversification of critical AI component suppliers and manufacturing locations to reduce single points of failure. Leverage GenAI for proactive risk management, predictive analytics, and scenario planning within national supply chains to enhance resilience aGenAInst geopolitical shocks. \cite{59}

    \item \textbf{Champion Data Sovereignty with Interoperability:} While pursuing data localization for national security and economic value, governments should also explore and invest in technologies like federated AI and distributed infrastructure. This approach can help balance data sovereignty with the need for data access and interoperability, preventing excessive fragmentation of the global digital economy. \cite{51}
\end{enumerate}

\subsection{For Industry}  
The private sector is both a driver and a steward of AI advancement. As industry actors integrate GenAI into operations, products, and services, they must adopt responsible AI practices, ensure algorithmic transparency, and contribute to open innovation ecosystems. Firms hold the capacity to shape norms, influence standards, and mitigate the productivity-inequality trade-off through inclusive design and deployment strategies.

\begin{enumerate}
    \item \textbf{Integrate GenAI Responsibly and Ethically:} Prioritize the ethical development and deployment of GenAI and autonomous systems. Implement internal governance frameworks that ensure transparency, accountability, and fairness, with clear human oversight mechanisms for high-risk applications. Conduct regular audits for bias and data leakage. \cite{82}

    \item \textbf{Invest in Workforce Upskilling and Reskilling:} Proactively invest in training programs to equip employees with the skills necessary to collaborate with GenAI and autonomous systems. This will mitigate job displacement risks and ensure that the productivity benefits of AI translate into widespread economic GenAIns, fostering job satisfaction and innovation. \cite{6}

    \item \textbf{Build Resilient AI Supply Chains:} Diversify AI component sourcing and consider regionalization or nearshoring strategies to reduce geopolitical vulnerabilities. Collaborate with governments and research institutions to develop secure and transparent AI supply chains, including robust cybersecurity measures aGenAInst AI-specific threats. \cite{68}

    \item \textbf{Foster Proprietary Insights Ecosystems:} While leveraging public and open-source AI models, companies should also focus on developing unique, proprietary datasets and insights. This will provide a competitive edge as external data becomes more accessible to all market participants. \cite{100}
\end{enumerate}

\subsection{For International Bodies}  
International institutions must rise to the challenge of coordinating global norms, managing cross-border risks, and ensuring that no nation or region is left behind in the AI transformation. In the absence of universal regulatory frameworks, platforms for multilateral dialogue, technical standard-setting, and ethical pluralism are crucial to preventing geopolitical fragmentation and enabling safe and equitable AI diffusion.

\begin{enumerate}
    \item \textbf{Bridge the Global Governance Gap:} Actively work to overcome geopolitical tensions that hinder global consensus on AI governance. Promote inclusive dialogues that involve all major AI stakeholders, including those currently excluded, to develop universally accepted norms and standards for AI development and deployment. \cite{16}

    \item \textbf{Harmonize AI Standards and Interoperability:} Facilitate the harmonization of AI standards across different jurisdictions to reduce fragmentation and improve interoperability between national-level regulations. Initiatives like an ``AI Standards Exchange'' could promote collaboration and reduce inconsistencies. \cite{27}

    \item \textbf{Address AI-Driven Global Inequalities:} Develop policies and initiatives to mitigate the risk of ``AI colonialism'' and widening global inequality. This includes promoting equitable access to AI infrastructure, talent development, and participation in global standard-setting for developing nations. \cite{27}

    \item \textbf{Strengthen International Norms for Dual-Use AI:} Urgently develop binding regulations and frameworks for the responsible military use of AI, particularly concerning autonomous weapon systems. Emphasize meaningful human control and adherence to international humanitarian law to prevent unintended escalation of conflicts. \cite{16}

    \item \textbf{Promote Multi-stakeholder Collaboration:} Continue to foster multi-stakeholder approaches to AI governance, ensuring the active involvement of governments, the private sector, civil society, and academic communities throughout the AI product lifecycle. This inclusive approach is crucial for developing robust, ethical, and widely accepted governance models. \cite{56}
\end{enumerate}

\section{Conclusion}
As GenAI continues to redefine the contours of Industry 5.0, it emerges as both a catalyst for innovation and a vector of geopolitical tension. The integration of autonomous systems across supply chains, industrial sectors, and governance regimes introduces transformative benefits but also amplifies systemic vulnerabilities and inequality. A fragmented global AI landscape—driven by divergent values, technological blocs, and regulatory asymmetries—threatens both interoperability and shared progress. Ensuring that AI remains aligned with human-centric principles demands dynamic, inclusive, and enforceable governance mechanisms. Strategic cooperation, ethical pluralism, and equitable access to AI infrastructure must underpin this next phase of development. Only through such a balanced approach can GenAI become a force for resilient, inclusive, and sustainable global advancement rather than a driver of division and control.

\section*{Conflict of Interest}
The authors declare that there are no conflicts of interest regarding the publication of this paper. All research procedures followed ethical guidelines, and the study was conducted with integrity and transparency. The authors have no financial, personal, or other relationships that could inappropriately influence or bias the content of this work.

\section*{Funding} 
No external funding was received for the conduct of this research or the preparation of this manuscript.

\section*{Use of Generative AI and AI-assisted Technologies}
During the preparation of this work, the author(s) utilized AI-based tools to assist with grammar correction and to improve writing clarity only. Following the use of these tools, the authors thoroughly reviewed and manually edited the content as necessary and accept full responsibility for the final version of the manuscript.





\bibliographystyle{ACM-Reference-Format}
\bibliography{work-main}



\end{document}